\documentclass[12pt,preprint]{aastex}
\begin{document}

\title{Disk Diffusion Propagation Model for the Outburst of
XTE\,J1118+480}

\author{Kent S. Wood\altaffilmark{1}, Lev Titarchuk\altaffilmark{1,2},
Paul S. Ray\altaffilmark{1}, Michael T. Wolff\altaffilmark{1},
Michael N. Lovellette\altaffilmark{1} \&
Reba M. Bandyopadhyay\altaffilmark{3}}

\altaffiltext{1}{E. O. Hulburt Center for Space Research, Naval
Research Laboratory, Washington, DC 20375}
\altaffiltext{2}{George Mason University/CEOSR, Fairfax, VA 22030}
\altaffiltext{3}{NRL/NRC Research Associate}

\shorttitle{Diffusion Model for Outburst of XTE\,J1118+480}
\shortauthors{Wood, Titarchuk, Ray, et al.}

\begin{abstract}
We present a linear diffusion model for the evolution of the
double-peaked outburst in the transient source XTE\,J1118+480.  The
model treats the two outbursts as episodic mass deposition at the
outer radius of the disk followed by evolution of disk structure
according to a diffusion process.  We demonstrate that light curves
with fast-rise, exponential decay profile are a general consequence of 
the diffusion process.  Deconvolution of the light curve
proves to be feasible and gives an input function specifying mass
deposition at the outer disk edge as well as the total mass of the
disk, both as functions of time.  The derived evolution of total disk
mass can be correlated with the observed evolution of the $\sim 0.1$
Hz QPO in the source reported in \citet{wrb+00}.
\end{abstract}


\section{Introduction}

X-ray novae (XN), also called Soft X-ray Transients (SXTs), are binary
X-ray sources that show outburst behavior \citep{csl97}.  The
prevailing theory for their transient behavior is based on the disk
instability model (DIM) first proposed to explain dwarf novae
outbursts (see the extensive review by \citealt{l01}).  Applying this
model to XN requires significant modifications due to X-ray
irradiation from the central disk and the disruption of the inner
disk, presumably by an advection dominated accretion flow (ADAF), in
quiescence.  The DIM with these modifications is shown to be quite
successful in modeling transient outbursts similar to the canonical XN
A0620$-$00 \citep{dhl01}.  However, it is also clear that some sources 
undergo outbursts that are very difficult to reconcile with the DIM.
In this paper we consider one such source, XTE\,J1118+480, and develop 
an alternative model for the outburst based on modulated mass transfer 
into the outer disk, rather than a sudden increase in viscosity in a
pre-existing disk as in the DIM.

We derive a general prescription for calculating X-ray outburst light
curves based on a diffusion propagation model in the disk.  We begin
with the same basic diffusion equations as \citet[][hereafter
BP81]{bp81} but we derive an analytical solution rather than solving
the equations numerically.  We derive a Green's function from the
first outburst and develop a deconvolution technique to calculate the
mass input rate as a function of time for the second outburst.  This
allows us to derive the time evolution of the mass of the disk for the
second outburst.  Based on the global oscillation model of
\citet[][hereafter TO00]{to00} we expect the QPO to correlate
inversely with the disk mass.  This provides us with at least one
observational test for the model.

It is possible that this model may be applicable to outbursts observed
in other sources as well.
\citet{bjf+00} point out similarities among five XNe:
{GRO\,J0422+32}, {XTE\,J1118+480},
{X1354-644}, {GRO\,J1719-24}, and
{GS\,2023+338}.  All five have had at least one outburst in
which the source remains in the low-hard state through the entire
episode.  In three instances a $\sim 0.1 - 0.5$ Hz QPO drifting
persistently upward in frequency is seen.  These sources may be good
candidates for future tests of this model; however, in this paper we
limit ourselves to XTE\,J1118+480 as an example.

In Section~\ref{sec-source} we review the properties of the source,
particularly the features of the 2000 outburst which are difficult to
account for in the DIM.  In Section~\ref{sec-diff} we describe the
main features of the diffusion model.  In Section~\ref{sec-solve} we
solve for the Green's funtion for the specific case where $\psi = 2$
and in the general case.  In Section~\ref{sec-inverse} we derive the
deconvolution procedure to get the accretion rate into the disk and
the time evolution of the disk mass during the outburst.  We apply the
full procedure to the data on XTE\,J1118+480 in
Section~\ref{sec-data}.  In Section~\ref{sec-observe} we show that
this model can explain the time evolution of the low-frequency
quasi-periodic oscillation (QPO) seen during most of the outburst.  We
conclude in Section~\ref{sec-discussion} with a discussion of the
successes and limitations of this model.

\section{The X-ray Nova XTE\,J1118+480}
\label{sec-source}

\begin{figure}
\plotone{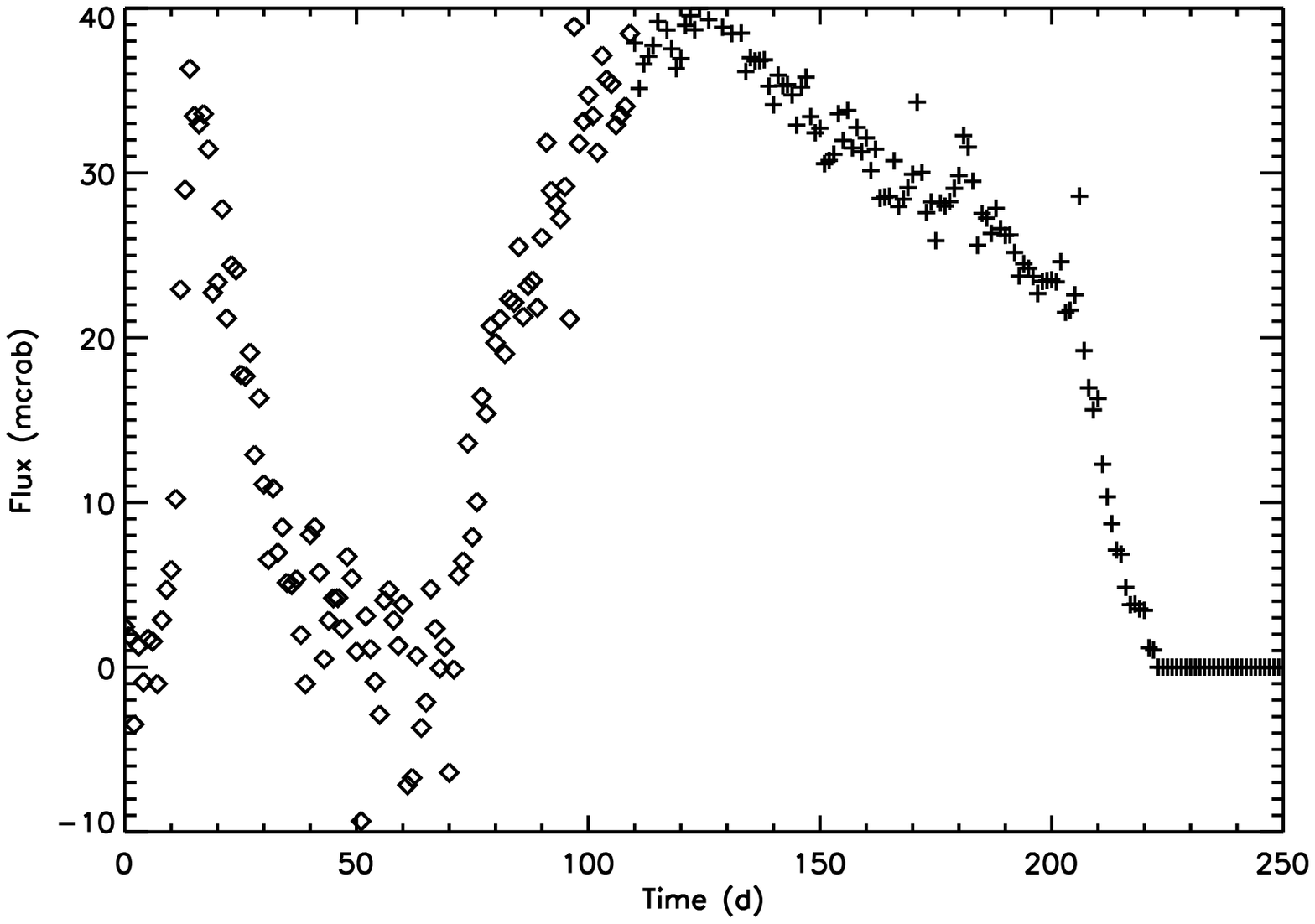}
\caption{Complete outburst light curve of XTE\,J1118+480.  The time
axis is labeled as days since 1999 December 23.  Diamonds are data from the
RXTE/ASM and crosses are from USA. The greater sensitivity of USA is
reflected in reduced scatter after day 110 where the USA data
begin. The data have been interpolated and re-gridded so as to give
one data point per day.}
\label{fig-data}
\end{figure}

{XTE\,J1118+480} is a black hole (BH) transient that was 
observed from radio to hard X-rays during 2000 January to August.
Optical observations after the source returned to quiescence confirmed
the black hole nature of the compact object by measuring a mass
function of $f(M) = 6.0 \pm 0.36 M_\sun$ \citep{mgc+01,wfs+01}.  This is
among the highest mass functions measured for any transient.  In
addition, the combination of the small distance to the source ($1.8 \pm
0.6$ kpc; \citealt{mgc+01}) and the high Galactic latitude ($b =
62$\degr) results in this source having the lowest reddening of any
XN.  The X-ray light curve of the source is peculiar, with an unusual
{\it double} outburst (see Figure~\ref{fig-data}). The first outburst
appears to have a fast rise followed by an exponential decay (FRED
morphology), but the second outburst is irregular with multiple
maxima.  The X-ray spectrum was essentially constant as an absorbed
power law with photon spectral index of 1.73 \citep{wrb+00}, typical of
the low/hard state of black holes.  A radio counterpart at 5 GHz has
been detected during outburst, although jets were not directly
resolved with MERLIN to a limit of $<65$ ($d$/kpc) AU \citep{fht+00}.

An earlier paper \citep{wrb+00} reported detailed evolution of the 0.1
Hz QPO, using X-ray data from the USA Experiment and {\em RXTE}.  The
QPO frequency increased monotonically from 0.07 to 0.15 Hz over $\sim
2$ months, failing to correlate with the X-ray luminosity which rises
and falls over the same interval.  In this paper, we explore using the
disk mass rather than usual quantities (such as luminosity or spectral
parameters) to correlate with the drifting QPO frequency.  The low
($\sim 0.1$ Hz) frequency suggests an origin at a large radius in the
disk because dynamical timescales scale roughly with the Kepler
frequency.  One theory that could explain such a low-frequency QPO is
the global oscillation model of TO00.  Their model describes a
large-scale coherent oscillation in the outer disk, linking QPO issues
to those of overall disk structure.  The model leads to a BH mass
estimate for XTE\,J1118+480 of $\sim 7 M_\sun$ based on the $\sim 0.1$
Hz QPO frequency, in agreement with recent optical data \citep{wfs+01}
which suggest a BH mass range of 6.0--7.7 $M_\sun$.  (We note that
there is a typographical error in equation (14) of TO00 which should have
$x_\mathrm{in}^{-4/5}$ instead of $x_\mathrm{in}^{-8/15}$.)  The same
QPO was seen in EUV and optical wavelengths \citep{hsp+00,hmh+00}. 

The double outburst profile of this source is difficult to account for
in the standard DIM.  If an outburst is triggered by a sudden increase
in disk viscosity and leads to the FRED-type outburst in 2000 January,
how can another outburst, with five times the fluence, occur so
shortly afterwards?  What mechanism turns off the first outburst?
And, why does the second outburst have such an irregular shape?  If
the instability is triggered at some maximum value of the disk mass,
and the amount of mass consumed in the first outburst was replenished
in $\sim 30$ days, then the mass consumed in the second outburst
should be replenished at about 0.5 years.  Thus, the DIM models would
have predicted that another outburst should have occurred by now. Our
alternative model is that the outbursts were fueled by sporadic mass
transfer from the companion rather than from a large amount of mass
stored in the disk during quiescence.  It is beyond the scope of the
current paper to develop the dynamics of the mass transfer which
accounts for this behavior.  It is instructive to look at the
energetics of the outburst.  The luminosity of the long outburst was
about $3 \times 10^{35}$ erg/s (for a distance of 1.8 kpc) which is
only $1.4 \times 10^{-4} L_\mathrm{Edd}$ (for an assumed black hole
mass of $7 M_\sun$), whereas the canonical DIM outbursts approach the
Eddington limit at their peak.

Mass transfer modulation models for XN outbursts have previously been
considered as viable alternatives to the DIM.  As the XTE\,J1118+480
outbursts are not easily reconciled with the standard DIM, it seems
reasonable to develop a different model for this source.  For example,
\citet{kul01} has suggested that the long outburst of XTE\,J1118+480
is a superoutburst of the type seen in the SU UMa class of cataclysmic
variables \citep{war95}, based on striking similarities in the X-ray
and optical lightcurves and the appearance of superhumps.  To account
for some of the unusual aspects of XTE\,J1118+480, here we investigate an
alternative model where the outburst is driven by mass transfer from
the companion and the diffusive propagation of material in the disk.

\section{Description of the Diffusion Model: Equations and Boundary
Conditions}
\label{sec-diff}

We now derive the main equation describing diffusive propagation of
matter in the disk.  Using the angular momentum balance
[e.g. \citealt{SS73}, see also \citealt{to99} hereafter
(TO99), eq. 4]
\begin{equation}
\dot M = 2\pi \frac{d(W_{r\varphi}R^2)}{d(\omega R^2)}
\end{equation}
and the continuity equation
\begin{equation}
{{\partial \Sigma}\over{\partial t}} = {1\over{2\pi R}} {{\partial \dot
M}\over{\partial R}} + Q(R,t) 
\end{equation}
where $\dot M (R,t)$ and $Q(R,t)$ are the mass accretion in the disk
and mass accretion input over the disk at the given radius $R$ and at
the given time $t$, and $W_{r\varphi}$ is the $r\varphi-$component of a
viscous stress tensor $W$. We can get a relation between
$W_{r\varphi}$ and the surface density $\Sigma (r,t)$ using equation
(5) in TO99 with an assumption that the rotational frequency is equal
to the Keplerian one,
$\omega=\omega_{\rm K}$:
\begin{equation}
W_{r\varphi} = -2\eta H R {{\partial \omega_{\rm K}}\over{\partial R}} 
= (3/2)(GM_x)^{1/2}\nu\Sigma R^{-3/2}
\end{equation}
where $H$ is a half-thickness of a disk, $G$ is the gravitational
constant, $M_x$ is the mass of the central object, $\eta=\rho\nu$ is a
viscosity, $\rho$ is the density, $\nu=l_tv_t/3$ is a kinematic
viscosity in the disk and $l_t$, $v_t$ are the turbulent scale and
velocity in the disk respectively. We also use the relation $\Sigma=2H\rho$.

A relation between $\dot M$ and $\Sigma$ can be obtained using
equation (1) and (3) with an assumption that $\omega=\omega_{\rm K}$
\begin{equation}
\dot M= 2\pi\times 3R^{1/2}
{\partial\over \partial R}[\nu \Sigma R^{1/2}].
\end{equation}
Substitution of $\dot M$ from this equation into equation (2) leads to
the diffusion equation
\begin{equation}
{{\partial \Sigma}\over{\partial t}} = {\bf\Lambda}_{R}\Sigma =
{3\over{R}} {\partial\over \partial R}
\left(R^{1/2}{\partial\over \partial R} [\nu \Sigma R^{1/2}]\right) + Q(R,t).
\end{equation}
Equation (5) is identical to equation (2.3) of BP81.  It is worth
noting that the derivation presented here and BP81 is general, using only
the mass and angular momentum conservations and an
assumption of Kepler rotational velocity in the disk. It does not
invoke specific assumptions regarding the structure of the
disk.  The viscosity value, $\eta=\rho \nu$ implemented for
calculation of $W_{r\varphi}$ in equation (3) are averaged over the
vertical size of the disk. The product of $2H\rho$ is in fact an
integral of the density, $\rho$, over the $z-$coordinate, yielding a
surface density $\Sigma$.

In our calculation we will assume that the mass accretion input
(presumably from the companion) occurs at the outer boundary of the disk
at $R=R_0$, namely
\begin{equation}
Q(R,t)= {{A(t)\delta(R-R_0)}\over{2\pi R}}.
\end{equation}
Thus, $A(t)$ is the mass input rate as a function of time at the outer 
edge of the disk.
Using equations (2) and (6) we derive the differential equation for the 
disk mass  $M_{d}=2\pi\int_{R_{in}}^{R_0}\Sigma RdR$:
\begin{equation}
{{\partial M_{d}}\over{\partial t}}= \dot M (R_0,t)- \dot M (R_{in}, t) +
A(t),
\end{equation}
where mass accretion $\dot M (R_{in}, t)$ at the inner disk boundary can be
calculated using equation (4). We assume the reflection boundary at the outer
edge of the disk; in other words, the net mass flux $\dot M(R_{0}, t)$ through 
the outer boundary is zero:  
\begin{equation}
\dot M (R_{0}, t)=0.  
\end{equation}
This is compatible with the mass input at the outer edge of the disk
from eqn. (6) because the mass input is considered to be added just
inside the outer radius $R_0$ where the reflection boundary condition
is imposed.
Integration of this equation allows us to get $M_d$ 
as a function of time $t$:
\begin{equation}
M_d(t)=\int_0^t [A(\tau)-\dot M (R_{in},\tau)]d\tau.
\end{equation}

We replace  the  function $\Sigma$ by the function
$y(R,t)=R^{1/2}\nu\Sigma$.
Using eq. (4) the mass accretion $\dot M$ is then expressed through $y$ 
as follows
\begin{equation}
\dot M= 2\pi\times 3R^{1/2}{{\partial y}\over {\partial R}}  
\end{equation}
and the diffusion equation (5) can be written in the form
\begin{equation}
{{\partial y}\over{\partial t}} =
{{3\nu(R)R^{1/2}}\over{R}} {\partial\over \partial R}
R^{1/2}{{\partial y}\over {\partial R}}  + 
{{\nu(R)A(t)\delta(R-R_0)}\over{2\pi R^{1/2}}}.
\end{equation}
We combine this equation with the boundary condition at the outer boundary
\begin{equation}
{{\partial y}\over {\partial R}}=0~~~~{\rm at}~~~R=R_0
\end{equation}
which follows from equations (8, 10).
We assume that at the inner boundary $R_{\rm in}\ll R_0$,
$\Sigma=0$, which is equivalent to
\begin{equation}
y=0~~~~~{\rm at}~~~~ R=R_{\rm in}.
\end{equation}
Introduction of a new variable $x=R^{1/2}$ reduces the main
equation (11) to:  
\begin{equation}
{{\partial y}\over{\partial t}} = {\bf\Lambda_x} y=
{{3\nu(x)}\over{4x^2}} {{\partial^2y}\over {\partial x^2}}
 + {{\nu(x)A(t)\delta(x-x_0)}\over{\pi}}
\end{equation}
with the boundary conditions
\begin{equation}
{{\partial y}\over{\partial x}}=0~~~~{\rm at}~~~~x=x_0,
\end{equation}
and 
\begin{equation}
y=0~~~~ {\rm for}~~~~x\rightarrow 0.
\end{equation}

One can check by direct substitution that the solution of
this boundary problem for the inhomogeneous equation (14) can presented
through the following convolution
\begin{equation}
y(x,t)=\int_0^tA(\tau)V(x,t-\tau)d\tau
\end{equation}
where  $V$ satisfies the homogeneous equation
\begin{equation}
{{\partial V}\over{\partial t}}={\bf\Lambda_x} V={{3\nu(x)}\over{4x^2}} 
{{\partial^2V}\over {\partial x^2}}
\end{equation}
with the initial condition
\begin{equation}
V(x)={{\nu(x)\delta(x-x_0)}\over{\pi}}~~~~{\rm at}~~~~t=\tau
\end{equation}
and boundary conditions (15-16).  Because our ultimate goal is to find
the X-ray luminosity, which is presumably related to the energy release
of the accreting matter at the inner disk edge, we will calculate $\dot
M$ at $R=R_{\rm in}$ (or at $x\rightarrow 0$).  We assume that the
mass accretion rate at the inner edge is converted with efficiency
$\varepsilon_{\rm eff}$ into the X-ray luminosity, i.e.
$L_x(t)=\varepsilon_{\rm eff}\dot M(0,t)$.  Thus
\begin{equation}
L_x(t)=\varepsilon_{\rm eff}\dot M(0,t)\propto \dot M(0,t)
=3\pi{{\partial y}\over{\partial x}}=
3\pi\int_0^t A(\tau){{\partial V(0, t-\tau)}\over{\partial x}}d\tau.
\end{equation}

\section{The analytical solution of the main problem: 
General and particular cases}
\label{sec-solve}

The solution $V(x,t)$ of equation (18) with the initial condition (19) at
$t=0$ and boundary conditions (15-16) can been presented using 
separation of variables as a series
\begin{equation}
V(x,t)=\displaystyle\sum_{n=1}^{\infty} 
e^{-\lambda_n^2 t}{{X_n(x)X_n(x_0)p(x_0)}\over{||X_n(x)||^2}}
{{\nu(x_0)}\over{\pi}},
\end{equation}
where $X_n(x)$ and $\lambda_n$ are eigenfunctions and eigenvalues which
can be found from the inhomogeneous ordinary differential
equation:
\begin{equation}
X_n^{\prime\prime}+\lambda_n^2p(x)X_n=0
\end{equation}
combined with the boundary conditions
\begin{equation}
X_n=0~~~~ {\rm for}~~~~x\rightarrow 0,
\end{equation}
\begin{equation}
{{d X_n}\over{d x}}=0~~~~{\rm at}~~~~x=x_0.
\end{equation}
$||X_n||$ is the norm of the eigenfunction, which is calculated through the 
integral (for example, see the derivation of this formula in \citealt{tmk97})
\begin{equation}
||X_n||^2=\int_0^{x_0}p(x)X_n^2(x)dx,
\end{equation}
where $p(x)=4x^2/3\nu(x)$ is the weight function. 

Here, we consider a general class of problems where
$\nu(x)=\nu_0 x^\psi$.  In the following sections, we solve the problem for one
specific value of $\psi$, then present the general solution.

\subsection{Case with $\psi=2$}

When the turbulent scale $l_t$ is proportional to the
characteristic scale $R$ and the turbulent velocity $v_t$ is
independent of radius, $\psi=2$ (i.e. $\nu=\nu_0 R$). In this case,
\begin{equation}
X_n(x) =\sin\left[{2\over{(3\nu_0)^{1/2}}}\lambda_{n}x\right],
\end{equation}
and the eigenvalues $\lambda_{n}$ are found from the equation
\begin{equation}
\cos\left[{2\over{(3\nu_0)^{1/2}}}\lambda_{n}x_0\right]=0
\end{equation}
where the solutions of this equation are
\begin{equation}
\lambda_n={{(3\nu_0)^{1/2}}\over 2}{{\pi(2n-1)}\over{2x_0}}.
\end{equation}

For this particular case the solution is expressed by harmonics
(eqs. [21] and [26])
\begin{equation}
V(x,t)={{2x_0\nu_0}\over{\pi}}\displaystyle\sum_{n=1}^{\infty}
\exp[-{{\pi^2(2n-1)^2t}/4t_0}]\sin[{{\pi(2n-1)x}/2x_0}]
\sin[{{(2n-1)\pi}/2}]
\end{equation}
where $t_0=4x_0^2/3\nu_0=4R_0^2/3\nu(R_0)$.  $t_0$ is the viscous
timescale and determines both the rise and fall time of the response function.

We define the Green's function $K(t)$ as the flux through the inner
edge of the disk at time $t$ from delta-function injection at the
outer edge at time $t=0$.  Thus, it is proportional to $\partial
V(0,t)/\partial x$:
\begin{equation}
K(t)\propto {{\partial V(0,t)}\over{\partial x}}=\nu_0\displaystyle \sum_{n=1}^{\infty}
\exp[-{{\pi^2(2n-1)^2t}/4t_0}](2n-1)\sin[{{\pi(2n-1)}/2}].
\end{equation}
By definition, $K(t)$ should be normalized to unity, which can be done
analytically in the case $\psi = 2$.  The integral of the series over
$t$ in this equation is
$\sum_{n=1}^{\infty}(-1)^{n-1}(2n-1)^{-1}=t_0/\pi$.

The asymptotic form of the series (29) for $t\ll 4t_0/\pi^2$ 
can be presented as the integral 
\begin{equation}
V(x,t)\simeq{{2x_0^2\nu_0}\over{\pi^2}}\int_0^\infty  
\exp[-{{\lambda^2 x_0^2t}/{t_0}}]\sin(\lambda x)
\sin(\lambda x_0) d\lambda,
\end{equation}
where $\lambda=\pi(2n-1)/2x_0$ and consequently $d\lambda=\pi/x_0$.
Transformation  of the product of $\sin(\lambda x)\sin(\lambda x_0)$ into
the difference $[\cos\lambda (x-x_0)-\cos\lambda (x+x_0)]/2$
followed by integration over $\lambda$ leads us to the formula
\begin{equation}
V(x,t)\simeq{{x_0^2\nu_0}\over{\pi^2}}{{\pi^{1/2}}\over{2x_0(t/t_0)^{1/2}}}
\left\{\exp[-(x-x_0)^2/4x_0^2(t/t_0)]-\exp[-(x+x_0)^2/4x_0^2(t/t_0)]\right\}.
\end{equation}
After differentiation  we have 
\begin{equation}
K(t)\propto {{\partial V(0,t)}\over{\partial x}}=\nu_0\times
{1\over{2[\pi(t/t_0)]^{3/2}}}\exp(-t_0/4t) \quad\textrm{for}\quad
t\ll 4t_0/\pi^2.
\end{equation}

At large times, we can use the series (30) to get
\begin{equation}
K(t)\propto {{\partial V(0,t)}\over{\partial x}}=\nu_0\times
\exp[-{{\pi^2t}/4t_0}]
\quad\textrm{for}\quad t\gg 4t_0/\pi^2.
\end{equation}

{} From a combination of the two asymptotic forms (eq. [31,34]) we
construct the simple analytical approximation of the Green's function.
This combination is a product of the combined exponentials with a
linear superposition of the factors before the exponents with weights
1 and 2 respectively:
\begin{equation}
K(t)\simeq C^{-1}\left[1+(t_0/\pi t)^{3/2}\right]
\exp[-(t_0/4t+\pi^2 t/4t_0)],
\end{equation}
where C is a normalization constant to make the integral of $K(t)$ equal
to unity.  We have chosen the weights of the two terms in this
approximation to make it more accurately reproduce the series result.

\subsection{General Case}

In this section we investigate the asymptotic behavior of the Green's
function for $t \ll t_0$ and $t \gg t_0$ in the case of arbitrary
kinematic viscosity as a function of $R$ (or $x = \sqrt{R}$).  We have 
already demonstrated for the $\psi = 2$ case that the asymptotic for
$t \ll t_0$ is determined by the contribution of terms of the series with 
$n \gg 1$ (see eq. [30]).  The calculation of the series is reduced to 
an integral over $\lambda$ (eq. [31]) where the contribution of terms
with $n$ of order few is negligible.  On the other hand, for $t \gg
t_0$ the term with $n=1$ is dominant and the contribution of terms
with $n>1$ is negligible.  It is worth noting that the integrated
function in equation (31) is a product of an exponent and harmonics.
The similar asymptotic form can be obtained in a general case of the
kinematic viscosity.

One can check by direct substitution that the
eigenfunctions $X_n(x)$ defined as nontrivial solutions of equations
(22--24) have the harmonic asymptotic form for large $n$
\begin{equation}
X_n(x) = p^{-1/4}(x)\cos[\lambda_n\mu(x)+\varphi]
\end{equation}
with $\mu^{\prime}(x) = d\mu/dx = \sqrt{p(x)}$ when  
\begin{equation}
\lambda_n \gg \displaystyle{\max_{x}\left\{|(p^{1/2})^{\prime}/p^{1/2}|,
|p^{\prime}/p|, [|(p^{-1/4})^{\prime\prime}/p|]^{1/2}\right\}}.
\end{equation}
 They form an orthogonal set of functions with the squared norm
\begin{equation}
||X_n||^2=\int_0^{x_0}p(x)X_n^2(x)dx=\int_0^{x_0}\mu^{\prime}(x)
\cos^2[\mu(x)+\varphi]dx\simeq \mu(x_0)/2.
\end{equation}

With an assumption that the kinematic viscosity $\nu$ is a power law function
of the radial coordinate $x$, namely $\nu=\nu_0x^{\psi}$, one can find
that $X_n$ is expressed through the Bessel functions
\begin{equation}
X_n=C_nx^{1/2}J_{1/(4-\psi)}[\mu(x)\lambda_n]
\end{equation}
with $\mu(x)=2/(4-\psi)\times(4/3\nu_0)^{1/2}x^{(4-\psi)/2}$.  Among
the two Bessel functions $J_{1/(4-\psi)}$ and $J_{-1/(4-\psi)}$, only
the former one allows $X_n(x)$ to satisfy the inner boundary condition
$X_n\to 0$ for $x \to 0$.  

We now investigate the behavior of the solution $X_n$ in the limits of 
large and small arguments $\mu(x)\lambda_n$.
Using the asymptotic form of the Bessel
function for large arguments \citep{as68}
\begin{equation}
X_n(x)=C_n x^{1/2}\left[{2\over{\pi
\mu(x)\lambda_n}}\right]^{1/2}\cos[\lambda_n\mu(x)-\pi(6-\psi)/4(4-\psi)].
\end{equation}
Comparing this formula with equation (36) we see that
$C_n=[\pi\lambda_n/(4-\psi)]^{1/2}$.  For small values of the argument, the
eigenfunction $X_n$ can be presented as series expansion around 0,
\begin{equation}
X_n(x)=\left[{{\pi\lambda_n}\over{(4-\psi)}}\right]^{1/2}
x^{1/2}\left[\mu(x)\lambda_n/2\right]^{1/(4-\psi)} \sum_{k=0}^{\infty}
{{[-(\lambda_n\mu(x))^2/4]^k}\over{k!\Gamma[(5-\psi)/(4-\psi)+k]}}.
\end{equation}
In fact, it follows from this equation that 
\begin{equation}
X_n(x) \propto x
\end{equation}
when $\lambda_n\mu(x)\ll 1$.

The eigenvalues $\lambda_n$ are determined by the equation 
\begin{equation}
{{dX_n}\over{dx}}=0 \quad{\rm for}\quad x=x_0.
\end{equation}
For small $n=1,2$ the eigenvalues are best determined using the series
(41), whereas for large $n$ the asymptotic form (40) is appropriate.
Fortunately, even $\lambda_1$ calculated using the two different
methods produces results which differ from each other by less than one
percent.  For example, the first three terms in series (41) provide an
accuracy of $\lambda_1$ better than 0.1\%. The use of these three
terms leads us to a quadratic equation for $u=(4-\psi)z^2/4$
where  $z=\lambda_n\mu(x_0)$:
\begin{equation}
u^2-2(5-\psi)u+2(5-\psi)=0.
\end{equation}
The first  root of this equation
\begin{equation}
u_1= 5-\psi-[(5-\psi)^{2}-2(5-\psi)]^{1/2}
\end{equation}
allows us to calculate the first eigenvalue.
For example, for $\psi=1$,  $z_1=1.25$,
whereas using the second method gives $z_1=1.18$ (see below).

For calculation of the eigenvalues using the eigenfunction form (40)
we have the transcendental equation:
\begin{equation}
\tan[z-\pi(6-\psi)/4(4-\psi)]=-(2-\psi)/2z(4-\psi).
\end{equation}

The solution of this equation can be written in an analytical form:
\begin{equation}
z \simeq [(n-1)\pi+\pi(6-\psi)/4(4-\psi)-\varepsilon_n],
\end{equation}
where 
\begin{equation}
\varepsilon_n={{2-\psi}
\over{2(4-\psi)[(n-1)\pi+\pi(6-\psi)/4(4-\psi)]}}.
\end{equation}
We apply equations (47-48) for calculations of the eigenvalues
for any $n$ because even for the first eigenvalue we have almost the same results
using these two methods (compare with Eq. 45).
 
The Green's function $K(t)$
can be written  using (41) as follows:
\begin{equation}
K(t) \propto {{\partial V}\over{\partial x}}(0,t)=
D^{\prime}\sum_{n=1}^{\infty}e^{-\lambda_n^2t}
\lambda^{\gamma}X_n(x_0),
\end{equation} 
where $\gamma=(6-\psi)/2(4-\psi)$ and
\begin{equation}
D^{\prime}={4\over3}{{x_0^2}\over{\pi^{1/2}}}{2\over{\mu(x_0)}}
\left({4\over{3\nu_0}}\right)^{1/2(4-\psi)}(4-\psi)^{-(6-\psi)/2(4-\psi)}
\Gamma^{-1}[(5-\psi)/(4-\psi)].
\end{equation}

Using the asymptotic form of $X_n$ at $x=x_0$ (see eq. 40) 
we can rewrite equation (49) as follows:
\begin{equation}
K(t) \propto {{\partial V}\over{\partial x}}(0,t)\simeq
 D\times \sum_{n=1}^{\infty}\lambda_n^{\gamma}\exp(-\lambda_n^2t)
 \cos(b\lambda_n-\pi\gamma/2),
 \end{equation}
where $b=t_0^{1/2}$
and $D=D^{\prime}(3\nu_0/4)^{1/4}x_0^{(\psi-2)/4}$.
As in the case with $\psi = 2$, the viscous time is $t_0$ which in the 
general case is
\begin{equation}
t_0=\mu^{2}(x_0)={4\over{3\nu(R_0)}}{4\over{(4-\psi)^2}}R_0^2.
\end{equation}

Thus for $t\gg\lambda_1^{-2}=(z_1^2/t_0)^{-1}$ or $t\gg t_0/z_1^2$
\begin{equation}
 {{\partial V}\over{\partial x}}(0,t)=D\times (z_1^2/t_0)^{\gamma/2}
 \cos[2(1-\gamma)/\pi(\gamma-3)]\exp(-z_1^2t/t_0).
\end{equation} 

For $t\ll t_0/z_1^2$ one should take into account all terms in the series
(eq. [51]).  As we have already demonstrated in \S 3, the summation of
such a series can be reduced to an integration over $\lambda$:
\begin{equation}
 {{\partial V}\over{\partial x}}(0,t)\simeq
 D\times {{t_0^{1/2}}\over{\pi}}
 \int_0^{\infty}\lambda^{\gamma}\exp(-\lambda^2t)\cos(b\lambda-\pi\gamma/2)
 d\lambda,
\end{equation}
where $d\lambda\simeq\pi/t_0^{1/2}$.

Now we proceed with calculation of the integral from equation (54),
\begin{equation}
I= \int_0^{\infty}\lambda^{\gamma}\exp(-\lambda^2t)\cos(b\lambda-\pi\gamma/2)
 d\lambda=\cos(\pi\gamma/2)I_c+\sin(\pi\gamma/2)I_s,
\end{equation}  
where 
\begin{equation}
I_c=\int_0^{\infty}\lambda^{\gamma}\exp(-\lambda^2t)\cos(b\lambda)
 d\lambda
\end{equation}
 and 
\begin{equation}
I_s= \int_0^{\infty}\lambda^{\gamma}\exp(-\lambda^2t)\sin(b\lambda)d\lambda.
\end{equation}
Expansion of  cosine and sine in series and integration
over $\lambda$ leads us to (see also Prudnikov, Bruchkov \& Marichev 1981)
\begin{equation}
I_c={1\over2}t^{-(\gamma+1)/2}\Gamma[(\gamma+1)/2]M[(\gamma+1)/2,1/2, -b^2/4t]
\end{equation}
and
\begin{equation}
I_s={b\over2}t^{-(\gamma+2)/2}\Gamma[(\gamma+2)/2]M[(\gamma+2)/2,3/2, -b^2/4t].
\end{equation}

\begin{figure}
\plotone{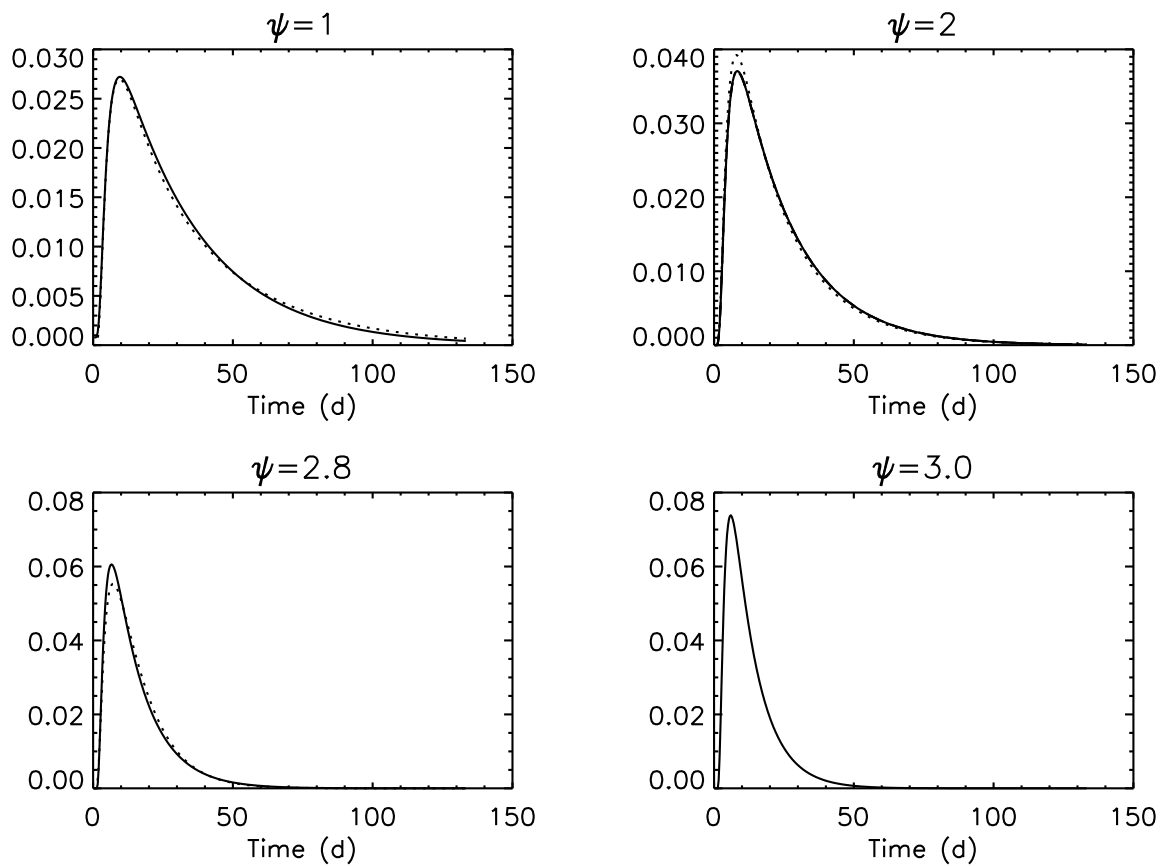}
\caption{Comparison of the Green's Function $K(t)$ for various values
of $\psi$.  The solid line is the series formulation (eq. [51]) and the dotted
line is the combination of asymptotic forms (eq. [62]).  There is no
dotted line in the bottom right panel because equation (62) breaks
down for $\psi \ge 3$.}
\label{fig-kgreen}
\end{figure}

The asymptotic form of the degenerate hypergeometric function 
$M(a,b,z)=e^zz^{a-b}\Gamma(b)/\Gamma(a)$ for large arguments 
$z$ (Abramowitz \& Stegun 1968) gives us the asymptotic form for integrals 
$I_c$ and $I_s$:
\begin{equation}
\{I_c, I_s\} \rightarrow
(\pi^{1/2}/2)(t_0/4)^{-(\gamma+1)/2}(t_0/4t)^{\gamma+1/2}\exp(-t_0/4t) 
\quad\textrm{for}\quad t \ll t_0/4.
\end{equation}
Finally we have
\begin{equation}
{{\partial V}\over{\partial x}}(0,t)\simeq
 D t_0^{-\gamma/2}\times \pi^{-1/2}(t_0/2t)^{\gamma+1/2}
 \exp(-t_0/4t)\sin(\pi\gamma/2+\pi/4) \quad\textrm{when}\quad t\ll t_0/4.
\end{equation}
We construct the analytical form of the Green's function $K(t)$ using 
the asymptotic forms eq. (53) for $t\gg t_0$ and eq. (61) for $t\ll
t_0$, similar to what was done for $\psi = 2$ in Section 4.1:
\begin{equation}
K(t)=C^{-1}\{2.5\pi^{-1/2}(t_0/2t)^{\gamma+1/2}
 \sin(\pi\gamma/2+\pi/4)+(z_1^2)^{\gamma/2}\cos[2(1-\gamma)/\pi(\gamma-3)]\}
 \exp(-t_0/4t-z_1^2t/t_0),
\end{equation}
where C is a normalization constant to make the integral of $K(t)$ equal
to unity. The factor 2.5 in the first term was empirically determined
to make this combination of asymptotic forms a better match to the series
formulation.  This form is primarily a computational
convenience.  Note that this form breaks down for $\psi \ge 3$ so we
restrict $\psi$ to be less than 3.  A graphical comparison of the
series and analytical forms for several values of $\psi$ can be found
in Figure~\ref{fig-kgreen}.

An important consequence of the previous derivation is that the FRED
outburst shape is a natural consequence of the diffusion model.  The
exponential decay is due to diffusion in a bounded
medium, while the fast rise is due to the fact that the probe (X-ray
emission from the inner edge of the disk) is located at a distance from
the source (accreting material added to the outer edge of the disk).
Thus we have derived the outburst shape for a large class of viscosity
structures when the viscosity is a function of only the radius.

\section{Solving The Inverse Problem}
\label{sec-inverse}

In this section, we demonstrate a deconvolution procedure to derive
$A(t)$, the mass input rate at the outer edge of the disk, and then
use $A(t)$ and $L_x(t)$ to derive the mass of the disk as a function
of time.  Now we proceed with general case of the input mass supply in
the disk $Q(R,t)$ which is proportional to $A(t)$ (see eq. 6).
Furthermore, we assume that the mass accretion rate at the inner disk
edge is converted with efficiency $\varepsilon_{\rm eff}$ into the
X-ray luminosity, i.e.  $L_{x}(t)= \varepsilon_{\rm eff}\dot M(t,
R_{in})$.  Thus using the light curve of $L_x(t)$ we can restore
information regarding the input function $A(t)$ by the deconvolution
of the following equation (see eq. 20):
\begin{equation}
L_x(t)=\varepsilon_{\rm eff}\dot M(0,t)=3\pi\varepsilon_{\rm eff}
\int_0^t A(\tau)K(t-\tau)d\tau. 
\end{equation}
To do this we also assume that the
variability time scale for A(t) is much longer than the viscous time
scale $t_0$. In other words the support T of the function $A(t)$ is much 
larger than that for $K(t)$. 
Then using the steepest descent method we get from equation (63) that
\begin{equation}
L_x(t)=\varepsilon_{\rm eff}\dot M(0,t)\simeq
3\pi \varepsilon_{\rm eff} A(\tau_{\rm max})\int_0^tK(t-\tau)d\tau
~~~~{\rm for}~~~t\leq T,
\end{equation}
or 
\begin{equation}
L_x(t)=\varepsilon\dot M(0,t)\simeq
3\pi \varepsilon_{\rm eff} A(\tau_{\rm max})\int_T^tK(t-\tau)d\tau
~~~~{\rm for}~~~t\geq T,
\end{equation}
where $\tau_{\rm max}$ is the time point for the maximum of
$K(t-\tau)$ in the integration interval.  The Green's function $K(t)$
has a sharp maximum at $t_{\ast}$ (which depends on $\psi$ and $t_0$)
and thus $\tau_{\rm max}= 0$ for $0\leq t \leq t_{\ast}$, $\tau_{\rm
max}=t-t_{\ast}$ for $t_{\ast}<t<T $, and $\tau_{\rm max}=T$ for $T<t<
T+t_{\ast}$.  With the steepest descent approximation of the
convolution (64-65) one can resolve that equation with respect to the
input function $A(t)$:
\begin{equation}
A(t-t_{\rm off})\propto
L_{x}(t)/\int_0^t K(t^{\prime})dt^{\prime}~~~~{\rm for}~~~t\leq T,
\end{equation}
where $t_{\rm off}={\rm min}(t, t_{\ast})$,
and 
\begin{equation}
A(t-t_{\rm off})\propto
L_{x}(t)/\int_{t-T}^t K(t^{\prime})dt^{\prime}~~~~{\rm for}~~~t\geq T,
\end{equation}
where $t_{\rm off}={\rm max}[(t-T), t_{\ast}]$.
Notice that $t_{\rm off}=t-\tau_{\rm max}$.

To calculate the mass of the disk as a function of time, we make the
assumption that the only terms that matter are mass input at the outer
boundary of the disk (proportional to $A(t)$) and the mass leaving the
disk into the black hole (proportional to $L_x$).  Hence the
instantaneous change of the disk mass $M_d(t)$ is determined by the
difference between the input function $A(t)$ and the output
function. If we assume that the output mass accretion rate $\dot
M(0,t)$ is proportional to the observed luminosity $L_x(t)$ in
accordance with equation (20) and the input function $A(t)$ is found as
a result of deconvolution with the operator $\bf K^{-1}$ defined by
eqs. (66-67), we can calculate the disk mass as a function of time as
follows: 
\begin{equation} M_d(t)\propto \int_0^t\{{\bf
K^{-1}}[L_x(t)]-L_x(t^{\prime})\}dt^{\prime}.
\end{equation} 
It is seen from eqs. (66-67) that $\bf K^{-1}$  is a composite operator 
which shifts and differentially expands the input function $L_x(t)$.

\section{Application to XTE\,J1118+480}
\label{sec-data}

We now apply this formalism to the 2000 January--August outburst of
XTE\,J1118+480. Figure~\ref{fig-data} shows a merged 2-12 keV light
curve from the ASM on RXTE and the USA Experiment on ARGOS, which
together cover the full double-peaked outburst.  Only the ASM viewed
the initial (FRED) outburst because the notification trigger did not
come until the second outburst. During the second outburst intensive
observation of the source commenced.  The greater sensitivity of USA
is better able to cover the final decline of the source than RXTE.

In order to proceed with the fitting, we need to select a value for
the $\psi$ parameter that defines the radial dependence of the
viscosity in the disk.  We have no strong physical reason to prefer
one value of $\psi$ over another, so we have chosen $\psi = 2.8$
primarily because it seems to work relatively well.  We found that
$\psi$ close to 3 worked better than $\psi$ of 1 or 2, so we chose a
$\psi$ slightly less than 3 so that equation (62) would still be valid.
In the future, as this method is applied to other sources it may
become clear that there is a preferred value for $\psi$ which is valid
for a class of sources.

\begin{figure}
\plotone{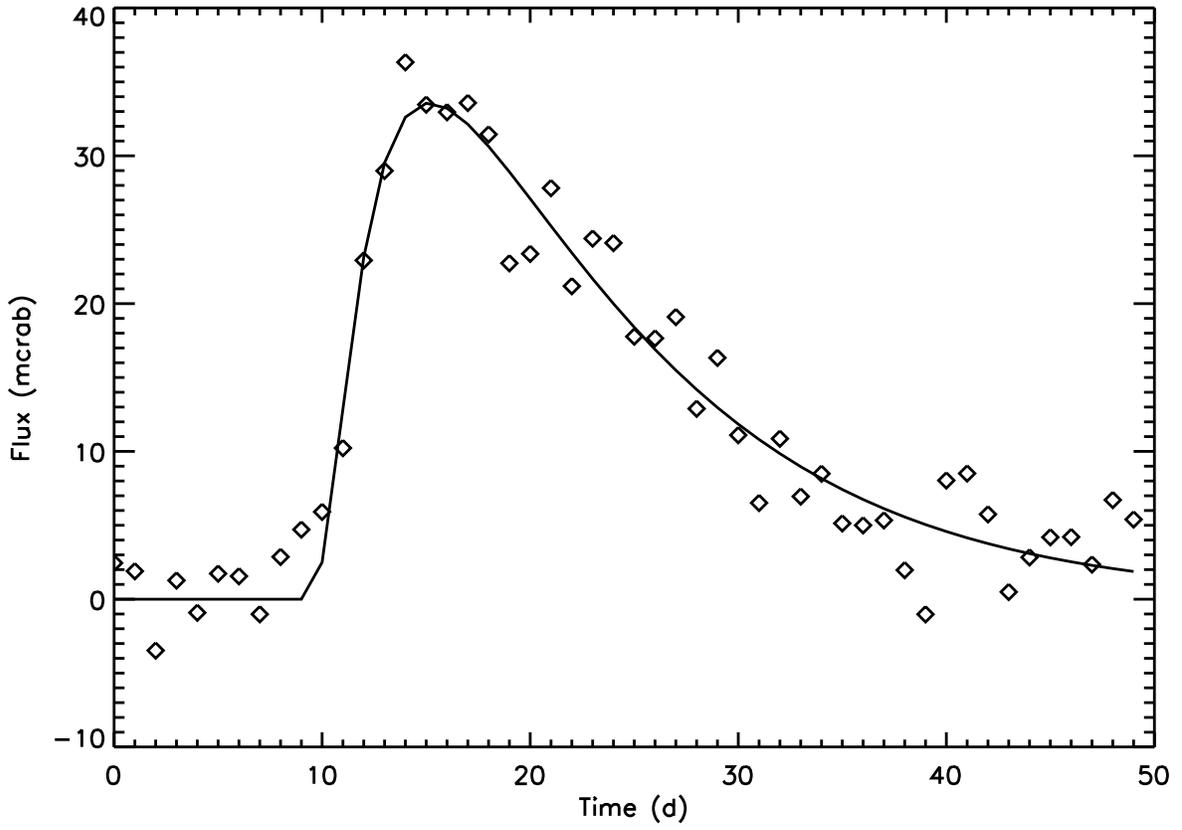}
\caption{Fit of Green's  function (eq. [62]) to initial outburst.  For
the fit, we have chosen $\psi = 2.8$.  Other than the trivial amplitude
and start time, there is only one free parameter, $t_0$.}
\label{fig-green}
\end{figure}

\begin{figure}
\plotone{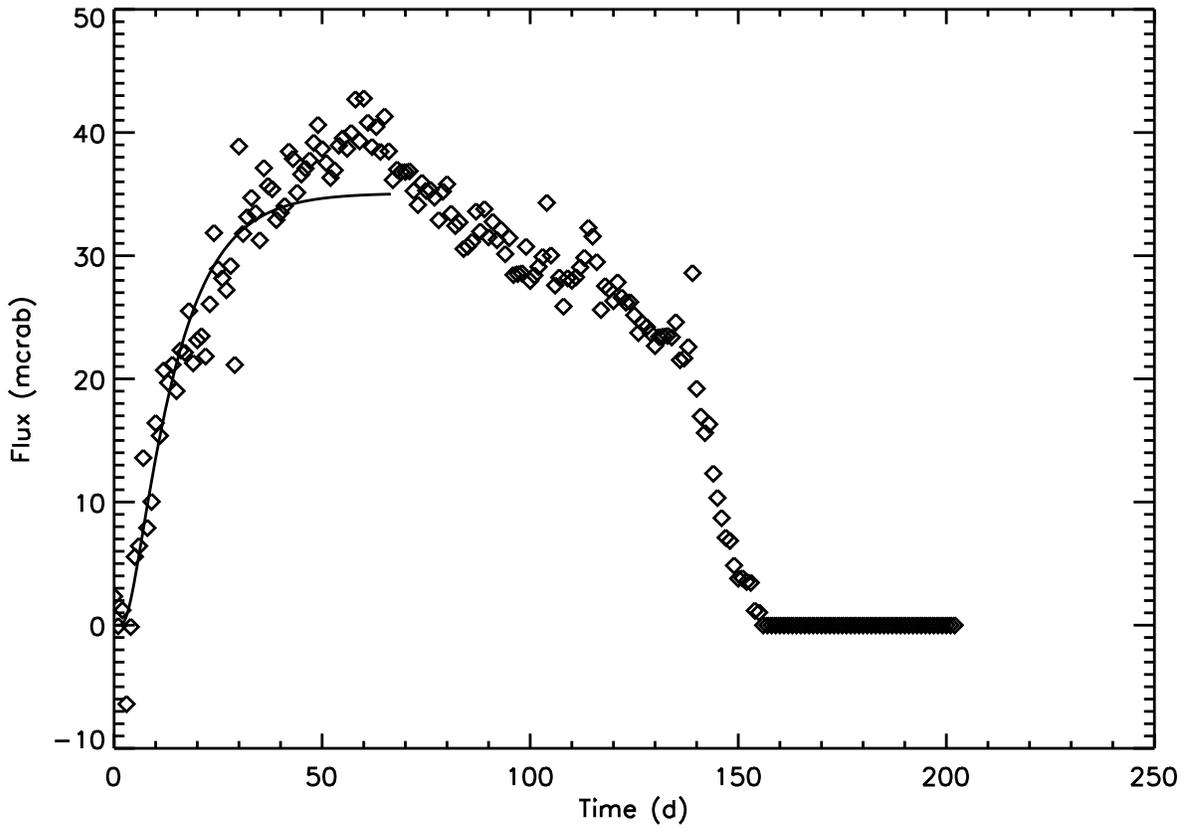}
\caption{Plot of complete second outburst showing integral of the best 
fit Green's function from the first outburst as a solid line.  This is 
the shape that would be expected for a step function turn on of the
mass transfer causing the second outburst.}
\label{fig-rise}
\end{figure}

Figure~\ref{fig-green} shows the fit of the Green's function
(eq. [62], with $\psi = 2.8$) to the first outburst.  The fit has only
two parameters: the viscous timescale $t_0$, and the initial time,
$t_{\rm start}$.  A normalization amplitude is also fitted in
Figure~\ref{fig-green}, but subsequently discarded when the function
is normalized to unity.  Hence the sole nontrivial parameter (beyond
our choice of $\psi$) in the fit is $t_0$.  The integral of this
measured function is found to approximate the shape of the initial
rise of the second outburst.  The integral has the shape expected from
a step function turn on; evidently the second outburst begins with
something close to a step function (see Figure~\ref{fig-rise}).  This
result is a consistency check of the convolution model for the second
outburst. In fact, for any input function $A(t)$ whose support is much
larger than that of the Green's function $K(t)$, the shape of the
convolution (eq. [63]) begins with the integral of $K(t)$ multiplied
by the initial value of $A(t)$.  This property of the convolution is a
useful tool to distinguish the convolution model of the outburst from
other possible models.

\begin{figure}
\plotone{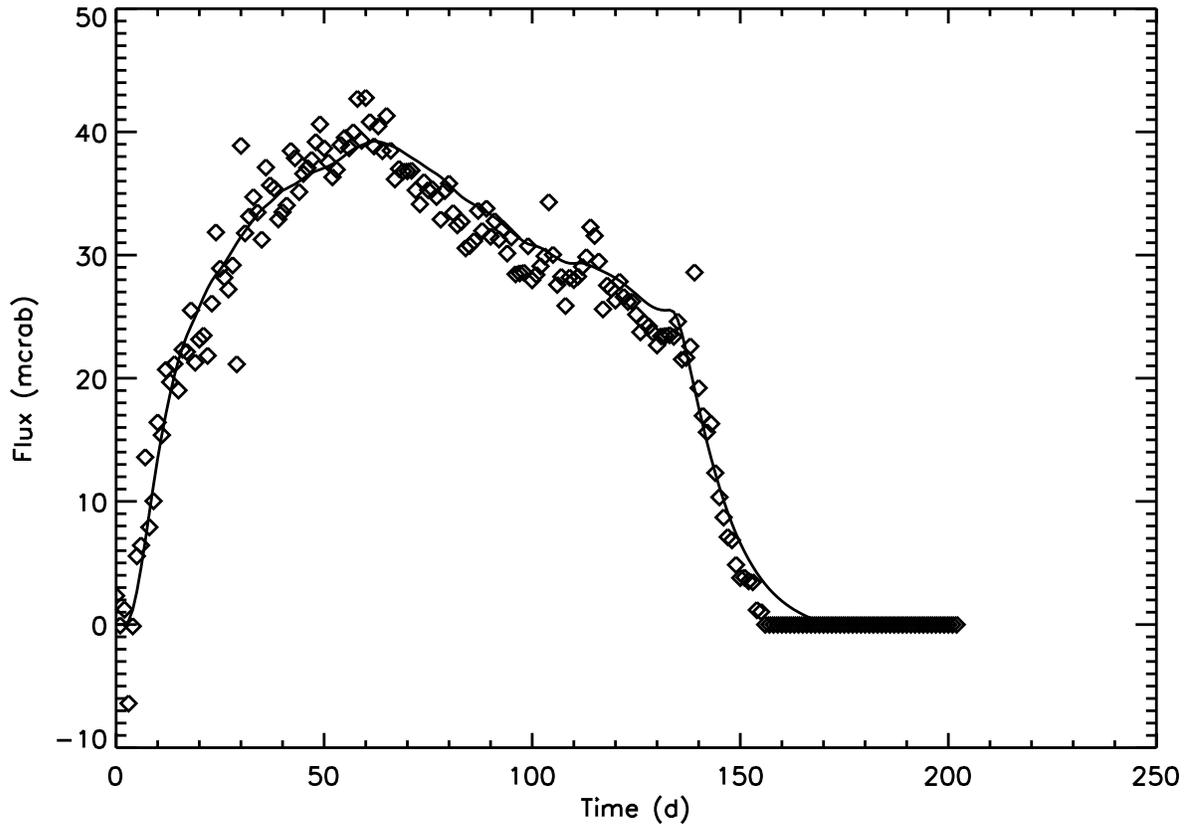}
\caption{Convolution of derived $A(t)$ with Green's function to get
estimated (smoothed) $L_x(t)$.  The solution process integrates and
smooths the input like a low pass filter.}
\label{fig-reconv}
\end{figure}

\begin{figure}
\plotone{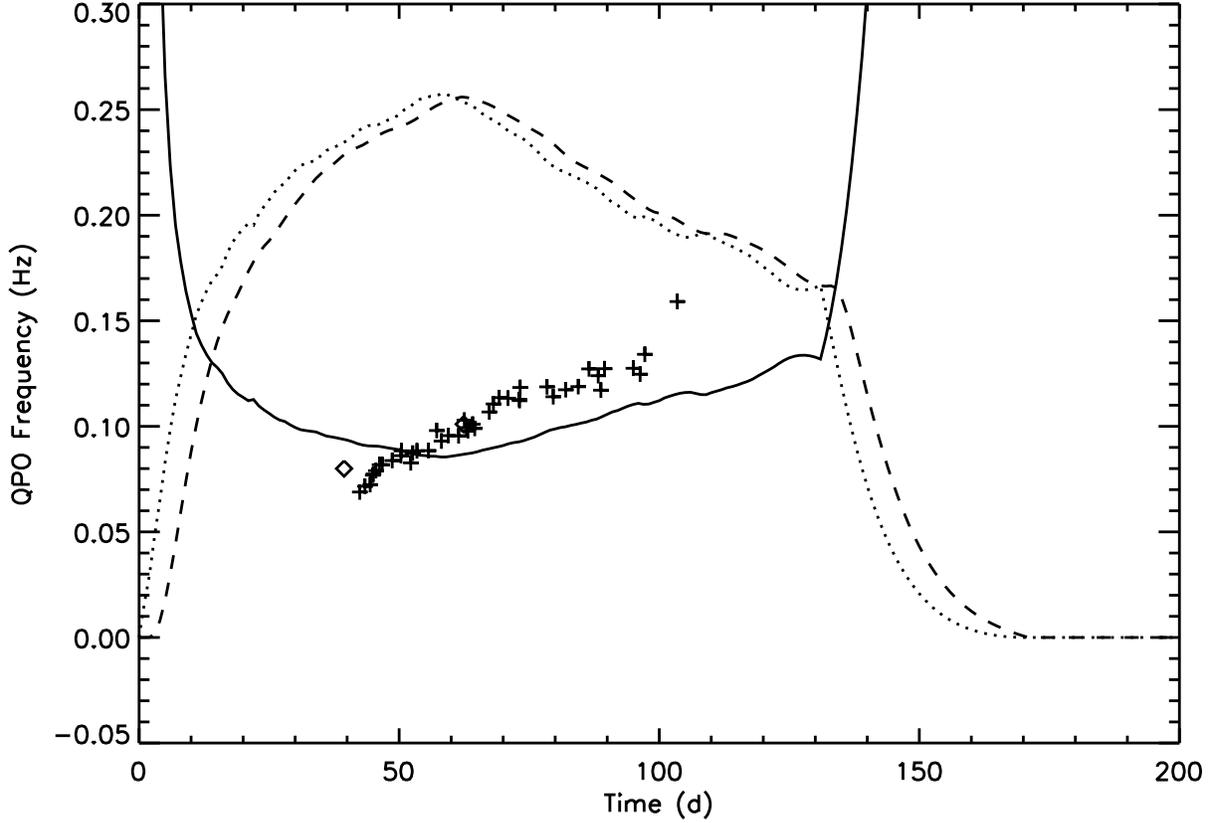}
\caption{Comparison of the QPO frequency evolution with the derived
mass of the disk.  The points are the QPO frequency according to the
left axis (crosses are X-ray determined QPO frequencies from
\citet{wrb+00} and diamonds are optically determined QPO frequencies
from \citet{hsp+00} and J. Patterson \& D. Skillman, private
communication).   The dotted line and the solid line are $M_d$ and
$M_d^{-1}$ respectively with arbitrary scale factors to put them on
the same scale as the QPO frequency.  The ripple in $M_d$ and
$M_d^{-1}$ derives from the scatter in the daily X-ray flux
measurements, propagated through the inversion process.  The dashed
line is the smoothed $L_x$ included to highlight the shift 
between $L_x$ and $M_d$.}
\label{fig-m}
\end{figure}

Equations (66-67) are used to extract the form of the input function
$A(t)$.  Figure~\ref{fig-reconv} verifies that this is the solution by
re-folding the solution with the Green's function to derive the X-ray
luminosity. The solution is smoothed because the convolution with the
Green's function acts as a low-pass filter.  Figure~\ref{fig-m}
displays the inferred mass $M_d(t)$ and its reciprocal $M_d^{-1}(t)$
on a plot of the QPO evolution.  The interesting result is that the
total disk mass is a nearly monotonically decreasing function during the
interval when the millihertz QPO frequency is increasing. Despite the
fact that the computed evolution of the disk mass looks very similar
to the evolution of the X-ray luminosity, the two curves are not
identical.  The disk mass evolution is obtained from the light curve
by the transformation which is effectively a shift transformation.

\section{Discussion and Interpretation of Observable Quantities}
\label{sec-observe}

\subsection{Evolution of the  disk mass and $\sim 0.1$ Hz QPO
frequency}

While Figure~\ref{fig-m} is not yet a complete model of the QPO
evolution, we argue that it represents a very substantial advance over
what was previously available.  To date there has been no model which
successfully explains the evolution of this low-frequency QPO.  If the
QPO is some large scale phenomenon in the disk then it should
correlate with large scale disk properties.  It fails completely to
correlate with X-ray luminosity in the sense that $L_x$ rises and
falls while QPO rises monotonically.  The function $M_d^{-1}(t)$,
while not perfect, moves the turnover earlier than that seen for $L_x$
and hence closer to that observed in the QPO frequency history.  We
believe that this result, combined with a physical model for the QPO
evolution represents a significant improvement in our understanding of
the observed QPO.  Note that our model included several simplifying
assumptions which when corrected may account for the remaining
differences.

We may now reconsider the global oscillation model (TO00,) in which a
monotonic dependence of $M_d^{-1}$ on the QPO frequency $\nu_0$ is
derived, giving $M^{-1}\propto \nu_0^2$ (see eqs. [4, 7] there).  The
observed dependence is flatter than the predicted one; that
discrepancy could be a result of our assumption of constant conversion
efficiency $\varepsilon_\mathrm{eff}$ of the mass accretion rate (in
fact, the surface density) at the inner disk edge into X-ray
luminosity.  It is possible that the inner radius of the disk
changes along with the outer disk radius, causing the variation of
$\varepsilon_\mathrm{eff}$.

The global oscillation is a vertical displacement of a large portion
of the disk.  The exact mechanism whereby this modulates X-ray (as
well as optical and EUV) emission to produce a QPO remains unclear.
One possibility is that the modulation is geometrical in nature, {\it e.g} 
from varying obscuration of the hot inner portions of the disk.
This would require the system to be at a fairly high inclination, as
found by \citet{wfs+01} for this source.  In addition,
for this idea be viable the corona would have to be somewhat limited in
extent.  Thus if the relevant disk oscillations were at $10^3 R_{\rm
S}$ (where $R_{\rm S}$ is the Schwarzschild radius) and the corona
were confined to less than $10^{2-3}R_{\rm S}$ the geometrical model
could work.  Other scenarios could include having the disk brightness
respond in some way to the global oscillation.

\subsection{Disk kinematic viscosity estimate and turbulent disk scale}

The best-fit parameter $t_0=48$ d in the Green's function from
equation (62) (see Fig.~\ref{fig-green}) allows us to derive the
kinematic viscosity value because $t_0=16 R_0^2/3\nu(R_0)$ from
equation (49).  Thus the kinematic viscosity is $\nu\sim
2\times10^{15}$ cm$^2$ s$^{-1}$ since $R_0\sim 4\times 10^{10}$ cm
(from eq. [10] in TO00 for $R_\mathrm{out}$ and using
$m=M/M_{\odot}=7$ and the orbital period 4.1 hours).  This estimate
leads to $l_t=6\times10^{9}$ cm if we assume that $v_t$ is of order of
a thermal velocity at the outer edge of the disk, i.e. $v_t\sim 10^6$
cm s$^{-1}$.

\section{Discussion and Conclusions}
\label{sec-discussion}

Here we have presented a detailed mathematical analysis of the diffusion
propagation model for X-ray outbursts.  We investigated the intrinsic
properties of the disk density evolution equation (5) in a general
case.  We have analyzed the diffusion models determined by
the disk kinematic viscosity dependence on the radius.  We have
demonstrated that the Green's (response) function is characterized by
the fast rising and exponential decay parts (FRED shape) and that it
is an intrinsic property for a wide class of diffusion models.
Similar results were obtained by Sunyaev \& Titarchuk
(1980) for photon diffusion models.  Particularly, the light curve
of the instant turn-on of a photon source placed in the center of
a spherical cloud (or the center disk plane) has a typical FRED type
shape (Figs. 1-2 there).

The success in fitting the diffusion model to the observed light curve
is very promising.  We have examined the case where
the viscosity is a power law function of position in the disk.  We
note that a dependence of viscosity on surface density is also
possible, but this will make the problem nonlinear (see Lyubarskii \&
Shakura 1987 and Lyubarskii 1997 for details).
 
We showed in \S \ref{sec-inverse} that the steepest descent method can
be employed to solve the inverse problem (a deconvolution) and
extract the input mass accretion function $A(t)$ using the observed
X-ray luminosity evolution.  Such a deconvolution is possible because
the inner mass accretion rate and the X-ray rate production are related
to each other and the timescale for viscous diffusion is short enough
compared to the timescale of the outburst.  Using the derived $A(t)$
as the source term and $L_x$ as a measure of the mass lost from the disk
into the black hole, we derived the mass of the disk as a function of
time.  This appears to inversely correlate with the observed QPO
frequency as would be expected from the global disk oscillation model, 
but the correlation is not perfect.

There are several factors that have been ignored for simplicity in
this initial treatment which could modify the derived disk mass as a
function of time: (1) a changing efficiency of the conversion of mass to
$L_x$ at the inner edge of the disk, (2) an effective viscous time
($t_0$) which varies with time, and (3) an additional sink for disk matter 
such as a jet.  We realize that the model is imperfect
but we think that the initial application is encouraging and that
further study and application to other sources is warranted.

We conclude by summarizing the main results of the diffusion model.
(1) The asymptotic form (eq. [62]) is characterized by the exponential
rise when the escape time is less than the diffusion timescale from
the location of mass injection. The same asymptotic form would be seen
for a semi-infinite medium and thus is determined by the escape
boundary condition (in our case at $R=R_{\rm in}$) and the location of
the mass source.  This asymptotic form is insensitive to the size of
the medium.  The second asymptotic form is characterized by an
exponential decay with the viscous timescale $t_0$ which is a
signature of the bounded medium (in our case, a disk) rather than any
specific model of the diffusion.  (2) The ``FRED'' shape of the first
outburst of XTE J1118+480 is in good agreement with the diffusion
Green's function form (eq. [62]), for which only one free parameter
$t_0$ is important.  The variation of the viscosity index
$\psi$ leads to a change of $t_0$ but the ``FRED'' shape remains.
 
This particular result is very important because it is closely related
to the intrinsic property of the diffusion Green's function. (3) The
value of the parameter $t_0$ has a physically plausible value for the
viscous timescale of the disk and ultimately can be applied to the
calculation of the disk turbulent scale using certain assumptions
regarding the disk size and the turbulent velocity (4). The rise of the
second outburst follows the form expected for a step function input,
the integral of the Green's function. It is worthwhile to emphasize
that the shape of this rise is precisely the integral of the Green's
function derived from the first outburst.  (5) The derived mass varies
approximately inversely with the QPO frequency.  (6) Similar X-ray
luminosity and QPO evolution would be expected in other sources with
similar behavior, and in fact is seen in {X1354$-$644} and
{GRO\,J1719$-$24} (Brocksopp et al. 2000).  In our model, the 
rising QPO frequency is a signature of the accretion disk mass decreasing 
as it disappears into the black hole.  If QPO frequency varies inversely 
with disk mass throughout the outburst there is a possibility that early
detection -- before the mass reaches its maximum -- might find the QPO 
when its frequency is still decreasing.  Both the relatively low brightness 
in the earliest stages of outbursts and the difficulties of prompt response 
would appear to have selected against detection of this effect to date.  

\acknowledgements

We would like to acknowledge the many helpful comments of an anonymous 
referee.
Basic research in X-ray Astronomy at the Naval Research Laboratory is
supported by the Office of Naval Research.  This work was performed
while RMB held a National Research Council Research Associateship
Award at NRL.  This paper made use of quick-look results provided by
the ASM/{\em RXTE} team (see {\tt http://xte.mit.edu}).

\clearpage

\end{document}